\documentclass[10pt,letterpaper]{article}
\usepackage[top=0.85in,left=1.5in,footskip=0.75in,marginparwidth=2in,textwidth=12in]{geometry}

\usepackage[utf8]{inputenc}

\usepackage{cite}

\usepackage{nameref,hyperref}

\usepackage[right]{lineno}

\usepackage{microtype}
\DisableLigatures[f]{encoding = *, family = * }

\usepackage{changepage}

\usepackage[aboveskip=1pt,labelfont=bf,labelsep=period,singlelinecheck=off]{caption}

\makeatletter
\renewcommand{\@biblabel}[1]{\quad#1.}
\makeatother

\usepackage{lastpage,fancyhdr,graphicx}
\usepackage{epstopdf}
\fancyheadoffset[L]{2.25in}
\fancyfootoffset[L]{2.25in}

\usepackage{color}

\definecolor{Gray}{gray}{.25}

\usepackage{graphicx}

\usepackage{sidecap}

\usepackage{wrapfig}
\usepackage[pscoord]{eso-pic}
\usepackage[fulladjust]{marginnote}
\reversemarginpar

\begin{document}
\vspace*{0.35in}

\begin{flushleft}
{\Large
\textbf\newline{Modeling communication processes in the human connectome through cooperative learning}
}
\newline
\\
Uttara Tipnis\textsuperscript{1,2},
Enrico Amico\textsuperscript{1,2},
Mario Ventresca\textsuperscript{1},
Joaqu\'{i}n Go\~{n}i\textsuperscript{1,2,3*},
\\
\bigskip
\bf{1} School of Industrial Engineering, Purdue University, West Lafayette, IN 47906
\\
\bf{2} Purdue Institute for Integrative Neuroscience, Purdue University, West Lafayette, IN 47906
\\
\bf{3} Weldon School of Biomedical Engineering, Purdue University, West Lafayette, IN 47906
\\
\bigskip
* jgonicor@purdue.edu

\end{flushleft}

\section*{Abstract}
Communication processes within the human brain at different cognitive states are neither well understood nor completely characterized. We assess communication processes in the human connectome using ant colony-inspired cooperative learning algorithm, starting from a source with no \textit{a priori} information about the network topology, and cooperatively searching for the target through a pheromone-inspired model. This framework relies on two parameters, namely \textit{pheromone perception} and \textit{edge perception}, to define the cognizance and subsequent behaviour of the ants on the network and, overall, the communication processes happening between source and target nodes. Simulations obtained through different configurations allow the identification of path-ensembles that are involved in the communication between node pairs. These path-ensembles may contain different number of paths depending on the perception parameters and the node pair. In order to assess the different communication regimes displayed on the simulations and their associations with functional connectivity, we introduce two network measurements, effective path-length and arrival rate. These communication features are tested as individual as well as combined predictors of functional connectivity during different tasks. Finally, different communication regimes are found in different specialized functional networks. Overall, this framework may be used as a test-bed for different communication regimes on top of an underlaying topology.


\section{Introduction}

Shortest paths in any system that can be described as a network, such as the human brain or roads in a state, are considered important as routing through them could minimize communication delays \cite{avena2018communication}. The concept of a shortest path is defined as the sequence of edges between a node pair in the network corresponding to the shortest distance between them. In the case of a road network, shortest paths may be used by a traveler who has knowledge of the network topology. On the other hand, in a biological network, such as the human brain, it is not assumed that the signal has knowledge of the network topology and thus does not necessarily take the shortest path. Even so, shortest paths are highly important in a biological network as they inform us of the most efficient routing. \\

In many natural systems, the concept of shortest paths can be relaxed to a natural selection of one more communication paths (potentially, an ensemble of paths) chosen through a collaborative effort, as in the case of foraging behaviour of ants. Deneubourg et al. \cite{deneubourg1990self} conducted experiments on Argentine ants (\textit{I. humilis}) to study their pheromone-driven foraging behaviour. In order to study how the indirect collaboration, also called \textit{stigmergy}, evolves over time from random exploration for food by an ant colony, the authors set up a two-bridge environment (from a network perspective, two nodes connected through two different edges, as in the case of multigraphs) that the ant colony explored. Both the edges (here representing possible paths) were of the exact same length and the passage of ants over the edges was observed over time. It was observed that at the beginning the individual ants randomly chose one of the two possible paths in search of food. However, as the pheromones dropped by the ants on their way started accumulating and affecting the environment over time, the ants eventually converged to using only one of the edges. As the experiment was run repeatedly, either of the paths emerged as the one on which the ants randomly converged. This is explained by the slight fluctuation in the number of ants taking each of the paths, increasing the pheromone concentration on that path. In the second part of the experiment, the ratio of lengths of the two bridges was 2:1. As a result, the ants always converged on the shortest path every time the experiment was run.\\

The foraging behaviour of many species of ants is driven by the indirect communication between them through chemicals, called \textit{pheromones}, as their visual faculties are not well developed \cite{goss1989self}. When a colony of ants starts searching for food, individual ants do so in a completely random and  uncooperative manner \cite{goss1989self}. As soon as an ant finds a food source, it takes some of it and carries it back to the nest and then starts again towards the food source. Thus, the ants keep moving back and forth between the nest and the food source until the food is completely depleted. During this journey, the ants leave a trail of pheromones on their path that other ants can smell. As the concentration of pheromones increases, ants gradually change their exploration behavior from unbiased random walks to an exploration biased by the concentration of pheromones. This is a critical characteristic of the ant-colony that allows the individual agents to initially explore many possible paths while subsequently converging to a potentially optimal path or ensemble of paths. This behavior of ant colonies seems to emerge in open two-dimensional spaces \cite{deneubourg1990self} as well as in constrained, network-like structures \cite{dorigo2000ant}\cite{deneubourg1990self}.\\

Ant colony optimization (ACO) algorithm takes its motivation from this indirectly collaborative behaviour of ants that allows them to find the shortest paths \cite{colornidorigo}. The ACO is typically used to find solutions to NP-hard problems that can be modeled as shortest path problems in a graph, e.g., travelling salesman problem (TSP), scheduling problems, assignment problems, amongst others.\\

It is well known that neuronal structure in the brain forms a complex structural network \cite{y1995histology}. This network is a static representation of white matter connections between brain regions. As such, it is a very slowly evolving topology. This structural network, called structural connectivity (SC), dictates how different parts of the brain communicate with each other, which is known as functional connectivity (FC). FC between two brain regions is the correlation between their fMRI time-series data. Thus, it is a fast evolving topology. Structural topology of the Human Connectome has been extensively assessed through measures that provide \textit{static} views of the underlying connectivity of the brain network, such as shortest path between regions, search information (SI), modularity, and degree distribution, amongst others. As these measures are static, they do not provide much explanation of how the SC and FC, which is a dynamic topology, might be related. Also, only a fraction of the edges in a network form the shortest paths. Thus, by assuming that communication in the brain takes place through shortest paths, one is essentially ignoring large parts of the network (discussed further in \textit{Section 3.1}). \\

In order to overcome the problems discussed above, De Vico Fallani et al. \cite{fallani2011multiple} and Avena-Koenigsberger et al. \cite{avena2017path} have proposed that the communication between brain regions does not take place through shortest paths. Instead, Avena-Koenigsberger et al. have suggested that communication between brain regions takes place through an ensemble of \textit{k}-shortest paths, while De Vico Fallani et al. have suggested investigating all possible paths between a pair of brain regions consisting of a certain number of edges. Thus, in both of these methods the paths that are investigated are pre-defined by the user.\\

In this paper, we propose a method to model signal propagation and communicability between brain regions through the use of an ant colony-inspired algorithm. We test this novel framework on the functional and structural data provided by the Human Connectome Project \cite{van2012human}\cite{van2013wu}. When exploring the network topology of the human structural connectome, the ants trace the ensemble of paths between each source-target pair of brain regions by traversing them probabilistically. By tuning two main communicability parameters related to the ant colony behavior (i.e., \textit{pheromone perception} and \textit{edge perception}), we investigate four different communication scenarios on SC: independent random walk, weakly coupled random walk, collaborative spreading preferentially along weak structural connections (\textit{side roads}), and collaborative spreading preferentially along strong structural connections (\textit{main roads}). For each scenario, we define two network measures extracted from the path ensembles traveled by the ants, namely \textit{effective path length} (EPL) and \textit{arrival rate} (AR). We show how these two node pair-wise measurements are good predictors of task-based FCs and partly of resting-state FCs, for different optimal choices of \textit{pheromone perception} and \textit{edge perception}. The predictive power of AR and EPL is even more noteworthy when considering communication scenarios within different functional subnetworks. We conclude by discussing the potentials of this new model for describing communication in large-scale brain networks and new directions for the investigation of communicability and signal propagation regimes in the human connectome, a new exciting avenue in brain connectomics.

\section{Methods}

\subsection{Human Connectome Project Data Processing}

The functional and structural dataset used in this work is from the Human Connectome Project (HCP, http://www.humanconnectome.org/), Release Q3. Below is the full description of the acquisition protocol and processing steps. We employed the Freesurfer parcellation into 164 brain regions \cite{fischl2004automatically}\cite{destrieux2010automatic}.

\subsubsection{HCP: Structural Data}

We used DWI runs from the same 100 unrelated subjects of the HCP 900 subjects data release \cite{van2012human}, \cite{van2013wu}. The diffusion acquisition protocol is covered in detail elsewhere \cite{glasser2013minimal}\cite{sotiropoulos2013advances}\cite{uugurbil2013pushing}. Below we mention the main characteristics. Very high-resolution acquisitions (1.25 mm isotropic) were obtained by using a Stejskal–Tanner (monopolar) \cite{stejskal1965spin} diffusion-encoding scheme. Sampling in q-space was performed by including 3 shells at b = 1000, 2000 and 3000 s/mm2. For each shell corresponding to 90 diffusion gradient directions and 5 b = 0’s acquired twice were obtained, with the phase encoding direction reversed for each pair (i.e., LR and RL pairs). \\

The HCP DWI data were processed following the MRtrix3 \cite{tournier2012mrtrix} guidelines (http://mrtrix.readthedocs.io/en/latest/ tutorials/hcp\_connectome.html), as done in recent paper \cite{amico2017hybrid}. In summary, we first generated a tissue-segmented image appropriate for anatomically constrained tractography (ACT \cite{smith2012anatomically}, MRtrix command 5ttgen); we then estimated the multi-shell multi-tissue response function \cite{christiaens2015global}(MRtrix command dwi2response msmt\_5tt) and performed the multi-shell, multi-tissue constrained spherical deconvolution \cite{jeurissen2014multi} (MRtrix dwi2fod msmt\_csd); afterwards, we generated the initial tractogram (MRtrix command tckgen, 10 million streamlines, maximum tract length = 250, FA cutoff = 0.06) and applied the successor of Spherical-deconvolution Informed Filtering of Tractograms (SIFT2, \cite{smith2015sift2}) methodology (MRtrix command tcksift2). Both SIFT \cite{smith2015sift2} and SIFT2 \cite{smith2013sift} methods provides more biologically meaningful estimates of structural connection density. SIFT2 allows for a more logically direct and computationally efficient solution to the streamlines connectivity quantification problem: by determining an appropriate cross-sectional area multiplier for each streamline rather than removing streamlines altogether, biologically accurate measures of fibre connectivity are obtained whilst making use of the complete streamlines reconstruction \cite{smith2015sift2}. Finally, we mapped the SIFT2 outputted streamlines onto the 164 chosen brain regions \cite{fischl2004automatically}, \cite{destrieux2010automatic} to produce a structural connectome (MRtrix command tck2connectome). Finally, a log10 transformation \cite{fornito2016fundamentals} was applied on the structural connectomes to better account for differences at different magnitudes. In consequence, SC values ranged between 0 and 5 on this dataset.

\subsubsection{HCP: Functional Data}

We used fMRI runs from the 100 unrelated subjects of the HCP 900 subjects data release \cite{van2012human}, \cite{van2013wu}. The fMRI resting-state runs (HCP filenames: rfMRI\_REST1 and rfMRI\_REST2) were acquired in separate sessions on two different days, with two different acquisitions (left to right or LR and right to left or RL) per day \cite{van2012human}, \cite{van2013wu}, \cite{glasser2013minimal}. The seven fMRI tasks were the following: gambling (tfMRI\_GAMBLING), relational (tfMRI\_RELATIONAL), social (tfMRI\_SOCIAL), working memory (tfMRI\_WM), motor (tfMRI\_MOTOR), language (tfMRI\_LANGUAGE, including both a story-listening and arithmetic task), and emotion (tfMRI\_EMOTION). The working memory, gambling, and motor task were acquired on the first day, and the other tasks were acquired on the second day \cite{glasser2013minimal}, \cite{barch2013function}. The HCP scanning protocol was approved by the local Institutional Review Board at Washington University in St. Louis. For all sessions, data from both the left-right (LR) and right-left (RL) phase-encoding runs were averaged to calculate connectivity matrices. Full details on the HCP dataset have been published previously \cite{glasser2013minimal}, \cite{barch2013function}, \cite{smith2013resting}.\\

The HCP functional preprocessing pipeline \cite{glasser2013minimal}, \cite{smith2013resting} was used for the employed dataset. This pipeline included artifact removal, motion correction and registration to standard space. Full details on the pipeline can be found in \cite{glasser2013minimal}, \cite{smith2013resting}.\\

For the resting-state fMRI data, we also added the following steps: global gray matter signal was regressed out of the voxel time courses \cite{power2014methods}; a bandpass first-order Butterworth filter in forward and reverse directions [0.001 Hz, 0.08 Hz] \cite{power2014methods} was applied (Matlab functions butter and filtfilt); the voxel time courses were z-scored and then averaged per brain region, excluding outlier time points outside of 3 standard deviation from the mean, using the workbench software \cite{marcus2011informatics} (workbench command - \texttt{cifti-parcellate}). For task fMRI data, we applied the same above mentioned steps, with a less restrictive range for the bandpass filter [0.001 Hz, 0.25 Hz].\\

Pearson correlation coefficients between pairs of nodal time courses were calculated (MATLAB command \texttt{corr}), resulting in a symmetric connectivity matrix for each fMRI session of each subject. Functional connectivity matrices from the left-right (LR) and right-left (RL) phase-encoding runs were averaged to improve signal-to-noise ratio. The functional connectomes were kept in its signed weighted form, hence neither thresholded nor binarized.\\

Finally, group average matrices were obtained from the resulting individual structural and functional connectivity (rest and 7 tasks) matrices. These were then grouped (rows and columns) according to the 7 cortical functional networks (FNs) as proposed by Yeo et al. \cite{yeo2011organization} based on resting state.

\subsection{Ant-colony Inspired Algorithm}

\begin{table*}
\centering
\begin{tabular}{| c | c | l | l |}
\hline
\textbf{Pheromone} & \textbf{Edge} & \textbf{Ant Colony Behaviour} & \textbf{Communication}\\
\textbf{Perception} & \textbf{Perception} & & \textbf{Regime} \\
\hline
$\alpha$ $>$ 1 & $\beta$ $>$ 1 & Highly collaborative and & Enforcing the use of\\
&& communication through & \textit{main roads}\\
&& most prominent edges only & \\
\hline
$\alpha$ $<$ 1 & $\beta$ $>$ 1 & No collaborative and & Weakly-coupled random\\
&& communication through & walkers with  preference \\
&& most prominent edges only & for \textit{main roads}\\
\hline
$\alpha$ $>$ 1 & $\beta$ $<$ 1 & Highly collaborative and & Enforcing the use of \\
&& extensive use of network & \textit{side roads}\\
\hline
$\alpha$ $<$ 1 & $\beta$ $<$ 1 & Weak collaboration and & Weakly-coupled random\\
&& extensive use of network & walkers with almost\\
&&& no preference for roads\\
\hline
$\alpha$ = 0 & $\beta$ = 0 & Not collaborative and & Independent random\\
&& extensive use of network & walkers with no\\
&&& preference for roads\\
\hline
\end{tabular}
\vspace{6pt}
\caption{Effect of the different configurations of \textit{pheromone} and \textit{edge perception} on the behaviour of the ant colony}
\label{table_alpha_beta}
\end{table*}

ACO is an optimisation algorithm used to solve NP-hard problems \cite{colornidorigo}, although in this paper the algorithm has been modified so that the goal is not optimisation anymore. Instead of finding the shortest path between brain regions, which can be done by better and much faster algorithms \cite{dijkstra1959note}\cite{floyd1962algorithm}, the modified ant colony algorithm is aimed at exploring the brain network and finding how hidden the brain regions are from each other. At the beginning, a source node and a target node are fixed, with the colony of ants located at the source and their goal is to find the target node. The network is unmarked by any pheromones in the beginning and so the ants start exploring the network in a random fashion. The probability of an ant taking a certain edge in the network is calculated as a transition probability and the ant randomly chooses between the neighbors of a node. The probability of an ant taking the edge \textit{ij} at time-step \textit{t} is calculated as follows:

\begin{equation} \label{eq1}
P_{ij}^{(t)} = \frac{(\tau_{ij}^\alpha)(\eta_{ij}^\beta)}{\sum\limits_{j} (\tau_{ij}^\alpha)(\eta_{ij}^\beta)}
\end{equation}

Where,\\

$\eta_{ij}$ = fiber density of edge \textit{ij}, i.e., the underlying brain structural connectivity matrix\\
      
$\beta$ = pheromone perception \\

$\tau_{ij}$ = amount of existing pheromone on edge \textit{ij} at time-step \textit{t}\\
      
$\alpha$ = edge perception\\

\begin{figure*}
	\includegraphics[width=\textwidth]{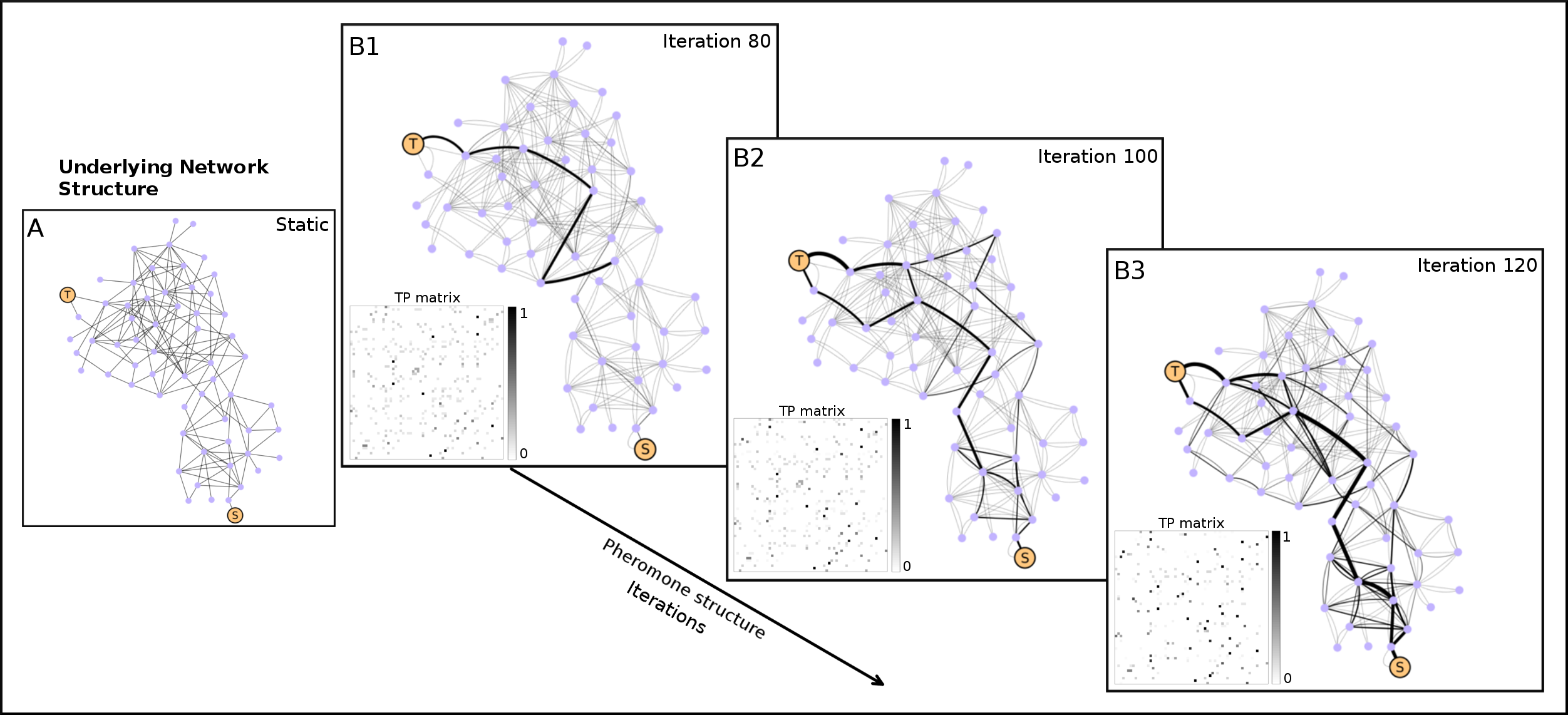}
    \caption{Ant' collective collaborative behaviour exhibited on the undirected Dolphin social network with 62 nodes. As ants start finding the target node, they deposit pheromones on some of the edges. Thus, the pheromone structure changes at every iteration. Shown here is the pheromone structure at three different iterations, along with the changing transition probability (TP) matrix.}
    \label{fig_dolphins}
\end{figure*}

Note that exponents $\alpha$ and $\beta$ characterize the individual perception of the ants. In particular, $\alpha$ represents the sensitivity to follow edges previously used by other ants, whereas $\beta$ represents how influential is the strength of the edge. Both exponents maybe sublinear, linear (equal to 1), and supralinear, and hence reflect the perception of the ants with respect to the underlying topology (SC network) and to the collaborative architecture through pheromones. Hence we will refer to $\alpha$ as \textit{pheromone (ph) perception} and to $\beta$ as \textit{edge perception}.\\

In \textit{Equation \ref{eq1}}, we are essentially calculating the transition probability of an ant going from node \textit{i} to node \textit{j} when accounting for two dimensions or layers of information (i.e. structural topology and pheromones). Thus, the denominator is a normalizing factor in order to keep the row-wise sum in the transition probability (TP) matrix equal to 1. As can be seen in \textit{Equation \ref{eq1}}, the calculation of the transition probability also involves the amount of existing pheromone on the edge \textit{ij}, which changes at every iteration. Hence, the TP matrix is dynamic across time. Thus, we cannot resort to analytical approaches, such as mean first passage time (MFPT) \cite{grinstead2012introduction}, in order to study this changing topology.\\

While running the simulations, the exponents in \textit{Equation \ref{eq1}}, \textit{pheromone perception} and \textit{edge perception}, are the variables that can be controlled within a certain range. Each separate combination of these variables characterizes a different collaborative spreading regime or configuration (expressed in terms of the two parameters) that can be associated with different task and resting state FCs. As $\eta$ is the underlying structural connectivity matrix in terms of fiber density, all the values in this matrix are lower than 1. Thus, when the value of \textit{edge perception} is below 1, the combined effect is to give more importance to all the values in the $\eta$ matrix. On the other hand, when the value of \textit{edge perception} is over 1, the combined effect is to shrink the values in $\eta$ even further, but the smaller values shrink more than the higher ones. For example, the effect of \textit{edge perception} = 2 on $\eta_{ij}$ = 0.05 and 0.5 is to make them 0.0025 and 0.25 respectively. Thus, the effect is similar to almost completely wiping out the weak connections in the structural connectivity matrix. The interaction between the pheromone matrix, $\tau$, and its exponent \textit{pheromone perception} is different as $\tau$ has entries that are above and below 1. The interactions between the \textit{ph} and \textit{edge} perception is summarized in \textit{Table \ref{table_alpha_beta}}. As the underlying structural connectivity matrix, $\eta$, remains static throughout the simulation and the pheromone matrix changes at every iteration, it can be thought of as a \textit{double-layered} network structure, where the interaction between the two layers of the network is regulated by \textit{Equation \ref{eq1}}. Even though the pheromone matrix changes at every step of the simulation, the way it changes depends on the SC, as the ants only take the existent edges. Thus, only the edges existent in the SC appear in the pheromone matrix.\\

The global behavior of the ants is a result of the individual dynamics. Thus, the simulation is set up in such a way that the ants are always in one of the three different states as follows:

\begin{itemize}

	\item \textbf{Explorer ant:} The ants all start at the source node acting as random walkers (\textit{explorer ants}) and do not deposit any pheromone on the edges. Hence, for each explorer ant, the step at time \textit{t} will be determined by the transition probability matrix at that time (see \textit{Equation \ref{eq1}}). The ants remain in this state until they find the target node.\\
    
    \item \textbf{Ant at target:} When an explorer ant reaches the target, it becomes an \textit{ant at target}. The ants are in this state only for one time step. At this step, the label of the ant changes from \textit{explorer ant} to \textit{homebound ant}.\\
    
    \item \textbf{Homebound ant:} The ants are in this state when they are coming back to the source node after visiting the target. In order to do so, they trace the same path back to the source that they took to get to the target. On their way, they deposit pheromones on the path as a signal to the other \textit{explorer ants}. This  effectively increases the probability of other ants taking the same edges to reach the target  in future. The amount of pheromone deposited by ants on their return journey is proportional to the length of the path taken. This pheromone update mechanism rewards the shorter paths over longer ones. When the \textit{homebound ant} makes its way back to the source node, it again starts a new journey as an \textit{explorer ant}.

\end{itemize}

As the algorithm is designed so that all the ants move one edge at every step of the simulation, it can provide us with the number of ants that have made it to the target at least once at every step.\\

In cases where no pheromone is deposited by the ants on the edges of the network, for any value of the \textit{edge perception} ($\beta$), the system behaves as a set of random walkers independently exploring the underlying SC. Such SC is in our case a representation of a complex network. This is not necessarily the case, and more simple models such as two dimensional lattices (representing a landscape) can be used as well. This would be equivalent to a coarse-grain scenario of the exploration of a two-dimensional open space, as widely assessed in foraging theory \cite{deneubourg1990self}.\\

In order to demonstrate the mechanism by which the ant colony algorithm cooperatively learns the topology over time, we executed it on a small toy network. \textit{Figure \ref{fig_dolphins}} shows the ant colony's collective behaviour exhibited on the undirected dolphin social network with 62 nodes \cite{lusseau2003bottlenose} for one source-target pair. As can be seen, the pheromone structure evolves with iterations as more and more ants reach the target and multiple paths emerge between the node pair. \\

This algorithm has provided a framework that allows us to study the spreading of information in the Human Connectome and the hiddenness of the brain regions from each other. The longer all the ants require to find the target, chances are that the more hidden is the target from the source.

\subsection{Ant Colony Simulations}

One run of the simulation consists of running the ant colony algorithm for every source-target pair in a 164-region parcellation for the structural connectivity (SC) of the group average of 100 unrelated HCP subjects. The algorithm runs in discrete time steps, i.e., the ants move one edge at a time. Before starting a simulation run, the following parameters can be controlled, along with the range of values that have been explored for each of them:

\begin{itemize}
\item $\alpha$ = Pheromone perception = [0.01,0.05,0.1,0.5:0.5:4]
\item $\beta$ = Edge perception = [0.1,0.5:0.5:4]
\item Amount of pheromone deposited by each ant = 1/$Length_{path}$
\item Number of ants in the colony = 200
\item Number of simulation steps = 1000
\item Number of simulation runs per configuration = 5
\end{itemize}

The simulation does not necessarily run for 1000 steps for every source-target pair as it stops when at least 95\% of the ants have made it to the target. This termination condition is added in order to optimize the running time of a full simulation run.\\

At every step of the simulation, the number of ants that have made it to the target at least once is saved, along with the paths that each individual ant took to reach there. Due to the effect of pheromones, convergence of ants is typically observed on multiple paths, although the path is discarded at the end of the simulation if it is used less than 10 times. The remaining paths are saved, along with the number of ants that have taken each path to reach the target. The data regarding the different paths taken by the ants is saved in order to study the backbone of the brain structural connectivity in terms of hiddenness of the target from the source and edge centrality that drives the spreading of ants in the network.\\

One important factor that should be remembered here is that the pheromone structure is not updated at every iteration in the entire network until all the ants have finished their move. Thus, an ant does not see another ants' freshly deposited pheromones until everyone has finished taking one step. This is to ensure that only the pheromones existing before the start of the iteration affect an ant's decision to take a certain edge.

\subsection{Network Analysis}

The data generated through the simulations contains very valuable information about the system, the evolving communicability, and the effects of collaborative spreading. Such information cannot be summarized by means of static measurements on the shortest-path, or even on a fixed set of paths. Hence, we developed two network-based measurements that account for the collective behavior of the ant colony as a single entity. One of the most important outputs of the ant colony simulations is the use of different paths by different number of ants to reach the target. The different paths taken by the ants between each source-target pair are kept track of, along with the number of times those paths are used. Thus, for each source-target pair, we may isolate the subnetwork within the underlying SC matrix that only includes edges belonging to the ensemble of paths used by the ants under each configuration.\\

One such network measure that we have defined is \textit{effective path length} (EPL), which is defined as the sum of the length of each path multiplied by the number of ants taking that path as a fraction of the total number of ants that have ever made it to the target. This can be represented mathematically as:

\begin{equation} \label{eq2}
EPL_{ij} = \frac{\sum\limits_{p=1}^{n_p}(L_p \times Traffic_p)}{\sum\limits_{p=1}^{n_p}Traffic_p}
\end{equation}

\noindent Where,\\
    
\noindent \textit{ij} = source-target pair for which effective path length is being calculated\\

\noindent \textit{$n_p$} = number of different paths taken by the ants\\
    
\noindent \textit{$L_p$} = length of path \textit{p} based on the fiber density\\
    
\noindent \textit{$Traffic_p$} = number of times the ants took path \textit{p} to reach the target\\

Note that \textit{Equation \ref{eq2}} is normalized by dividing by the sum of all the ants that have reached the target by taking the saved paths. Thus, the contribution of paths taken less than 10 times by the ants to the EPL would be negligibly small. Hence, ignoring those paths does not significantly affect the calculation of EPL. A high EPL reflects that communication through the path ensemble involves longer paths, whereas a low EPL suggests the involvement of shorter paths. \\

Another network measurement that we have defined is \textit{arrival rate} (AR). For every source-target pair, this is the ratio of the number of arrivals to the maximum number of arrivals that could have taken place (defined as an ant exclusively taking the shortest path back and forth between source and target). Mathematically, this can be shown as:

\begin{equation}
AR_{ij} = log_{10} \Bigg( \frac{2 \times Arrivals_{ij} \times SPL_{ij}}{ numAnts \times (iter_{arrival} + SPL_{ij})}\Bigg)
\end{equation}

\noindent Where,\\

\noindent \textit{$Arrivals_{ij}$} = number of times ants have successfully reached target \textit{j} from source \textit{i} by using any path in the path ensemble\\

\noindent \textit{$numAnts$} = number of ants used in the simulation = 200\\

\noindent \textit{$SPL_{ij}$} = number of edges in the shortest path length between \textit{i} and \textit{j}\\

\noindent \textit{$iter_{arrival}$} = iteration number when at least 95\% of the ants reach the target\\

AR represents the $log_{10}$ transformation of the percentage of arrivals (hence, ranging between 0 and 1) to the target with respect to the maximum number of arrivals that could possibly happen as bounded by communication solely happening through shortest paths. A high value of AR represents that the communication between a node pair is efficient, whereas a low value indicates that it is inefficient with few arrivals to the target.

\subsection{Null Models Based on Structural Connectivity}

\begin{figure}
	\includegraphics[width=\textwidth]{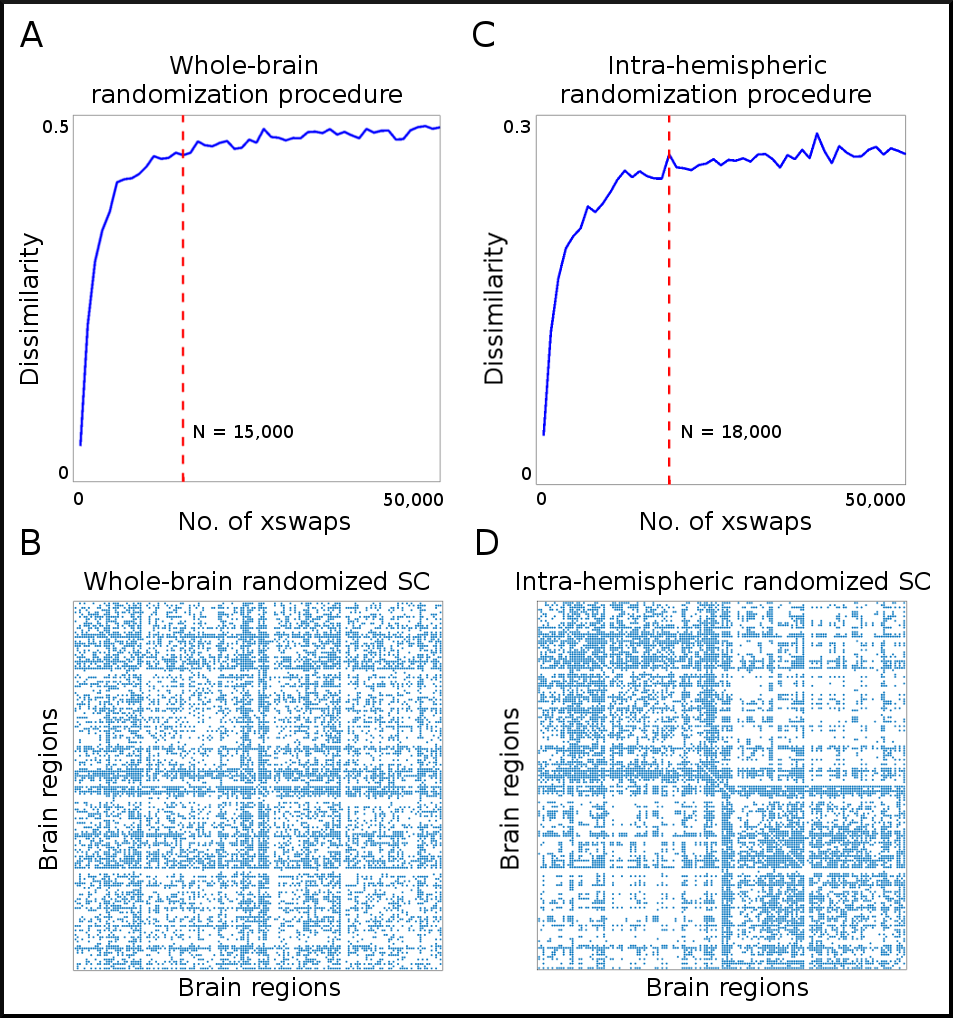}
    \caption{Iterative randomization procedure on SC network. Dissimilarity as a function of the number of xswaps for whole-brain (\textit{A}) and intra-hemispheric randomizations (\textit{C}). Adjacency matrices for final randomized networks are shown for whole-brain (\textit{B}) and intra-hemispheric (\textit{D}).}
    \label{fig_randomize}
\end{figure}

We tested two different null models based on randomizations of SC \cite{maslov2002specificity}. In order to do so, we evaluated the FC predictive power of the ant colony-derived measurements on two different randomized topologies. Note that the randomization procedure used for both null models have density, degree-distribution, and degree-sequence as topological invariants. 

\begin{itemize}
\item \textbf{Whole-brain Randomization:}

The entire SC network was randomized using  the \textit{xswap} method \cite{maslov2002specificity} \cite{hanhijarvi2009randomization}. This iterative procedure was performed until full randomization was achieved. This was evaluated by measuring dissimilarity between the original and the increasingly randomized network until a plateau was reached. Dissimilarity reflects the percentage of entries that are different between a network and a reference network. Note that upper boundaries of dissimilarity are dependent on the density of the network, and hence a perfect dissimilarity (value of 1) may only be reached in networks where density is 50\%. \textit{Figure \ref{fig_randomize}.A} and \textit{B} show dissimilarity as a function of number of xswaps and the randomized network, respectively. \textit{Figure \ref{fig_randomize}.A} shows that dissimilarity reaches a plateau around 0.5 after 15,000 xswaps. Hence, we chose this configuration for the analysis of the null model based on whole-brain randomization. This whole-brain randomized network will henceforth be referred to as $SC_{whole}^{rand}$.\\

\item \textbf{Intra-hemispheric Randomization:}

The second null model used introduced another invariant by always preserving the inter-hemispheric connections, i.e., neither deleting any existing nor adding any new edges when performing xswaps. Analogously to the procedure performed in the first null model, it was found that dissimilarity reached a plateau around 0.3 after 18,000 xswaps. \textit{Figure \ref{fig_randomize}.C} and \textit{D} are the dissimilarity as a function of the number of xswaps and the randomized network, respectively. This intra-hemispheric randomized network will henceforth be referred to as $SC_{intra}^{rand}$.

\end{itemize}

We run the simulations as explained in \textit{Section 2.3} on the two null models explained above. The same network analysis as in \textit{Section 2.4} was conducted out on the data generated by these simulations. The \textit{Results} section discusses the results of these null model-based simulations in detail.

\section{Results}

As described earlier, 100 unrelated subjects in the HCP \cite{van2012human}\cite{van2013wu} dataset were used to construct the group average structural and functional connectomes. The ant colony simulations were run on the group average structural connectome. The data that is saved from these simulations consists of the different paths taken by the ants between each node pair, along with the number of ants taking each of the paths. In order to characterize different aspects of communication for different $\alpha$-$\beta$ configurations, we calculated EPL and AR for each source-target pair (see \textit{Section 2.4} for details). These two measures were then associated, individually and together (in a multilinear regression), with the group average task-based and resting-state functional connectivity patterns, estimated as per the protocol described in \textit{Section 2.1}. This section reports the results obtained from the path ensembles and the associations with the FCs.

\subsection{Evaluation of Path Ensembles and Betweenness Centrality}

\begin{figure*}
	\includegraphics[width=\textwidth]{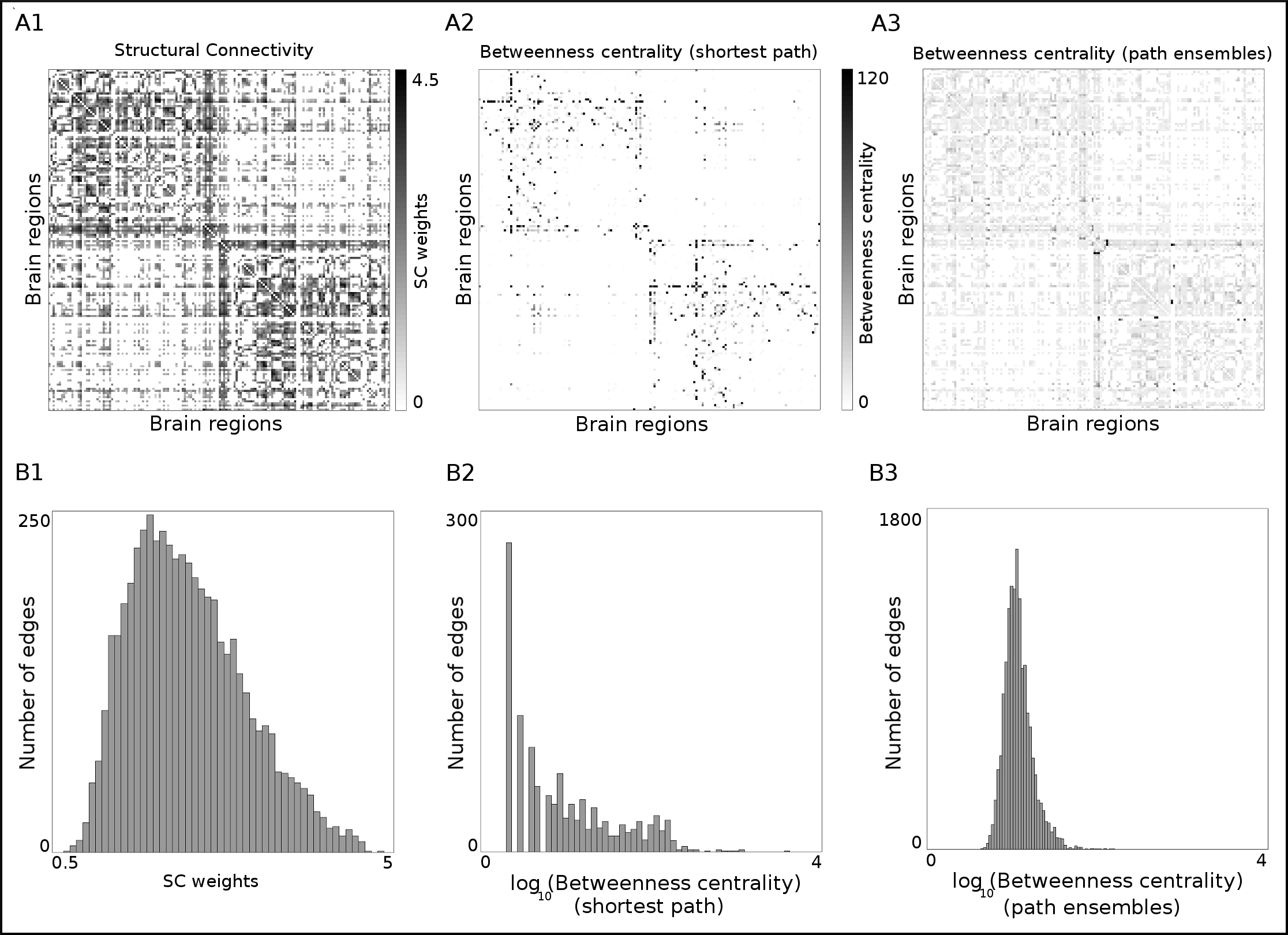}
    \caption{\textit{A1} is the group average weighted structural connectivity (SC). \textit{A2} is the edgewise shortest-path betweenness centrality on SC. Note that 23\% of the edges participate in at least one shortest path. \textit{A3} is the equivalent of the shortest-path betweenness centrality in the path ensembles obtained through the ant colony algorithm (results correspond to the configuration $\alpha$ = 2 and $\beta$ = 0.1). 100\% of the edges participate in at least one path ensemble. The \textit{B1, B2}, and \textit{B3} are the corresponding histograms for each measurement.}
    \label{fig_centrality}
\end{figure*}

As discussed in \textit{Section 2}, running the ant colony inspired algorithm allows us to identify the ensemble of paths most widely used by the ants for source-target pair in the brain structural network. \textit{Figure \ref{fig_centrality}.A3} shows the edgewise betweenness centrality when it is measured only on shortest paths and on the path ensembles (as obtained from the simulations). \textit{Figures \ref{fig_centrality}.A2} and \textit{A3} show that only 23\% of the edges participate in shortest paths, whereas 100\% of the edges participate in the path ensembles.\\

Comparing \textit{Figures \ref{fig_centrality}.A2} and \textit{\ref{fig_centrality}.A3}, we can see that the ants often traverse edges that are not part of any shortest path. Also notice that the path ensemble betweenness centrality is more visually comparable to the underlying structural connectivity on which the ant colony inspired algorithm is run. \textit{Figures \ref{fig_centrality}.B1}, \textit{\ref{fig_centrality}.B2}, and \textit{\ref{fig_centrality}.B3} show the histograms of the values for each of the three corresponding plots. Notice that the distribution of the path ensemble-based betweenness centrality is more balanced and displays a log-normal behaviour.

\subsection{Associations Between Functional Connectivity and Path Ensemble-Derived Measures}

\begin{figure*}
\includegraphics[width=\textwidth]{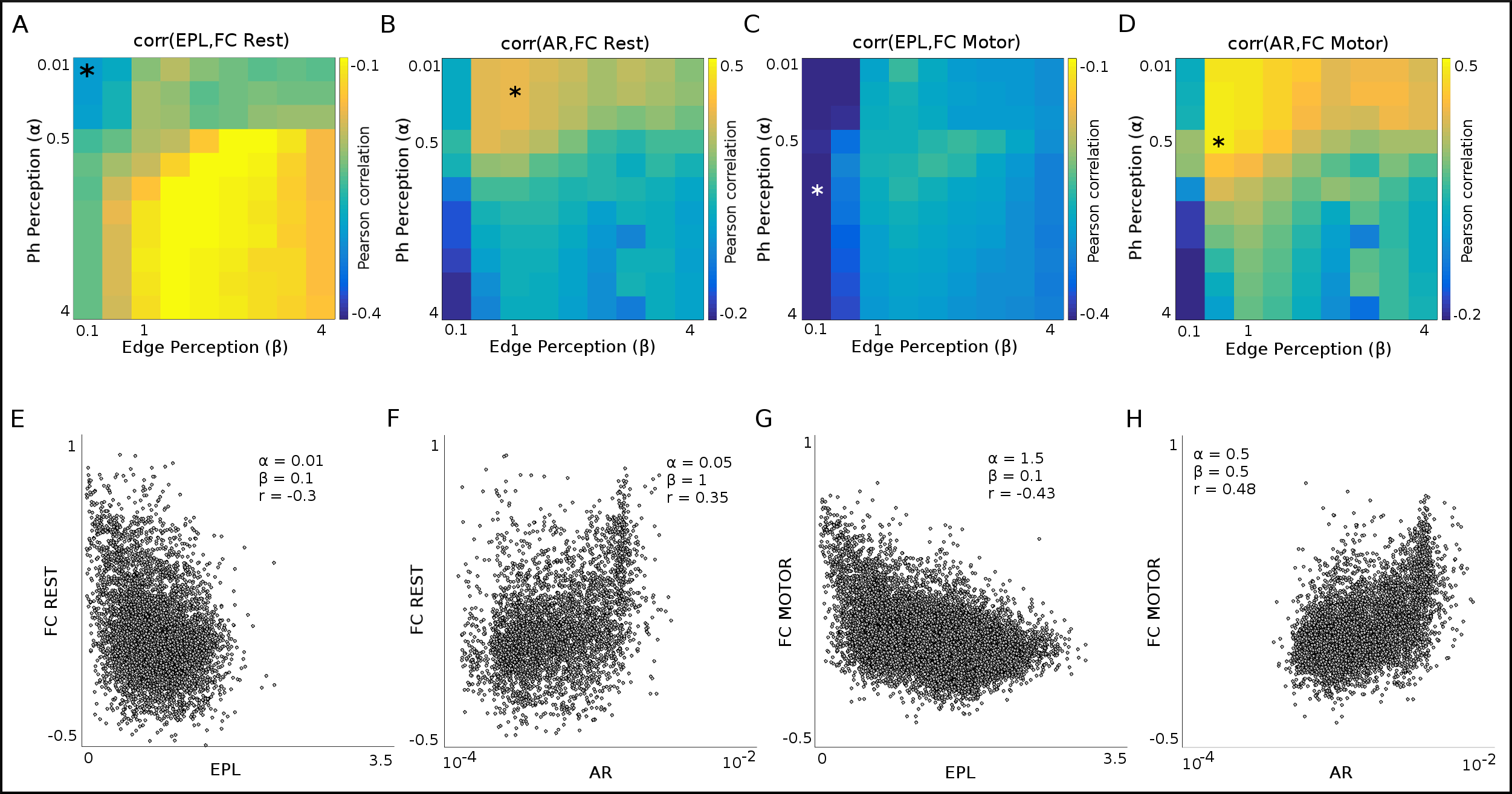}
\caption{For every configuration of the ant colony, Effective Path Length and Arrival Rate are calculated for every source-target pair. The Pearson correlations between these measures and different task-based and resting state functional connectivities are calculated. \textit{A} and \textit{B} show the correlations of resting state FC with EPL and AR, while \textit{C} and \textit{D} show these correlations with Motor FC. The * in each of these matrices shows the configuration for which the correlation is highest. \textit{E, F, G}, and \textit{H} are the scatter plots between the FCs and EPL and AR for the configuration with the highest correlation.}
\label{fig_EPL_AD}
\end{figure*}

The two network measures defined in \textit{Section 2.2}, effective path length (EPL) and arrival rate (AR), were calculated for every source-target pair in every configuration of \textit{ph} and \textit{edge perception}. Then, EPL and AR for each configuration were correlated with each task-based and resting state functional connectivity (FCs). \textit{Figure \ref{fig_EPL_AD}.A-D} show the correlation values of EPL and AR with resting state and motor-task FC for all configurations of \textit{ph} ($\alpha$) and \textit{edge perception} ($\beta$). \textit{Figure \ref{fig_EPL_AD}.E-H} illustrate the scatter plots of EPL and AR with the resting state and motor-task FCs corresponding to the configurations with highest correlations in \textit{Figure \ref{fig_EPL_AD}.A-D}. Note that for resting state FC, the highest correlation with EPL is achieved at ($\alpha$ = 0.01, $\beta$ = 0.1) and with AR at ($\alpha$ = 0.05, $\beta$ = 1). Analogously, for motor-task FC, the highest correlations are attained at ($\alpha$ = 1.5, $\beta$ = 0.1) and ($\alpha$ = 0.5, $\beta$ = 0.5) with EPL and AR respectively.\\

\begin{figure}
	\includegraphics[width=\textwidth]{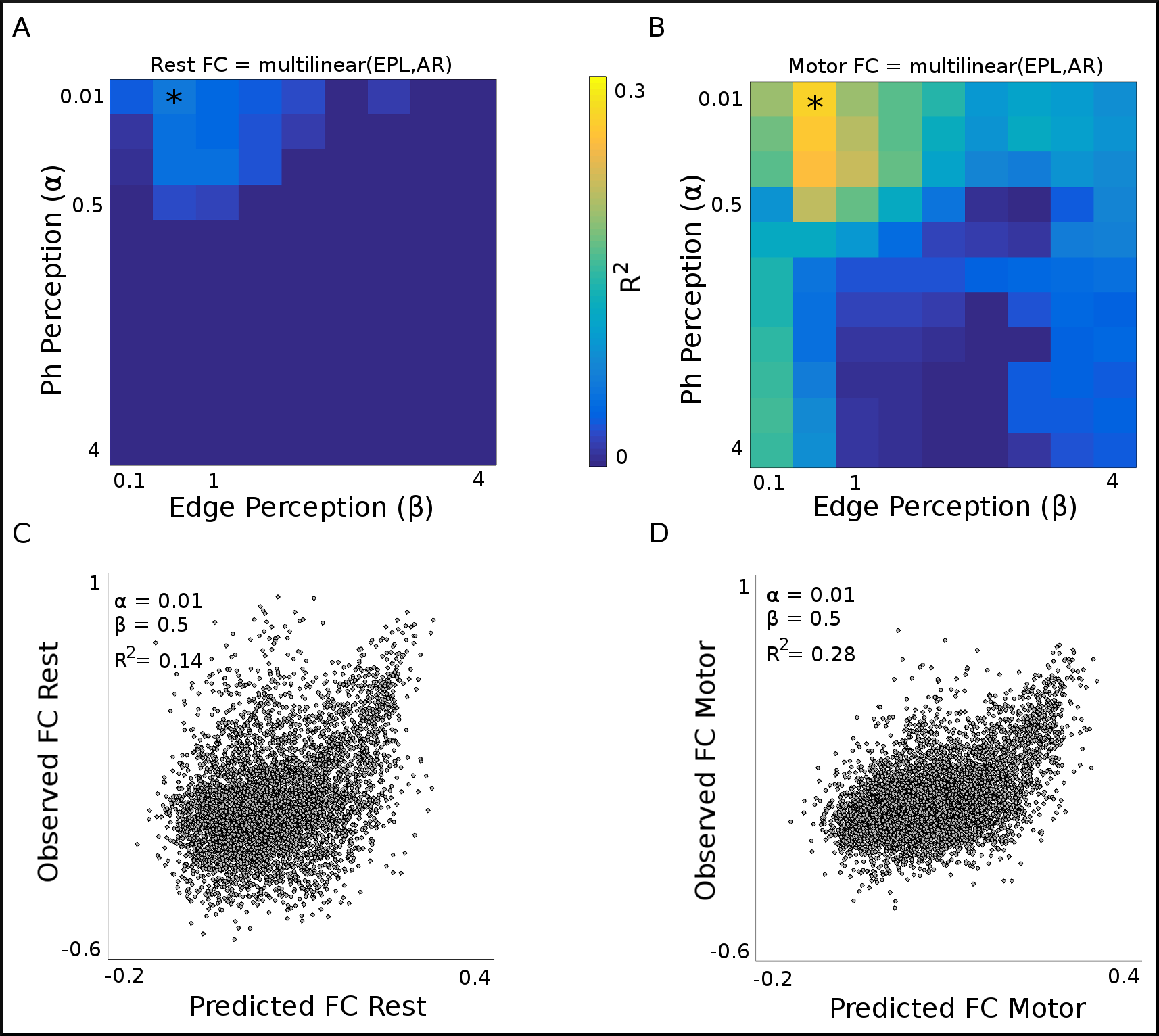}
    \caption{For every configuration of the ant colony, multilinear regression is carried out with EPL and AR as the predictor variables and the different task-based and resting state FCs as the predicted variable. \textit{A} and \textit{B} show the $R^2$ values of the regression for the different configurations, while the * highlights  the configuration for which the $R^2$ is highest. \textit{C} and \textit{D} are the scatter plots between the predicted and observed FCs for resting state and Motor task respectively.}
    \label{fig_regstats}
\end{figure}

\begin{table}
\centering
\begin{tabular}{| l | c | c | c | c |}
\hline
& \multicolumn{4}{c|}{\textbf{Path Length Measurements}} \tabularnewline
\hline
\textbf{Task} & \textbf{SC} & $\textbf{SC}_{\textbf{\scriptsize intra}}^{\textbf{\scriptsize rand}}$ & $\textbf{SC}_{\textbf{\scriptsize whole}}^{\textbf{\scriptsize rand}}$ & \textbf{SC}\\
&  \textbf{(EPL)} & \textbf{(EPL)} & \textbf{(EPL)}  & \textbf{(SPL)}\\
\hline
REST & -0.30 & -0.07 & 0.03 & -0.14 \\
\hline
LANGUAGE & -0.40 & -0.16 & 0.02 & -0.27 \\
\hline
EMOTION & -0.38 & -0.15 & 0.01 & -0.28 \\
\hline
GAMBLING & -0.37 & -0.15 & 0.01 & -0.27 \\
\hline
MOTOR & -0.43 & -0.20 & -0.02 & -0.33 \\
\hline
RELATIONAL & -0.36 & -0.15 & 0.01 & -0.27 \\
\hline
SOCIAL & -0.36 & -0.13 & 0.01 & -0.24\\
\hline
WM & -0.41 & -0.17 & 0.01  & -0.31 \\
\hline
\end{tabular}
\vspace{6pt}
\caption{Pearson correlation coefficients between path length measurements on SC and task-based FCs. Path length measurements were obtained from SC, and from two null models derived from SC, namely $SC_{intra}^{rand}$, and $SC_{whole}^{rand}$. EPL is the effective path length, calculated from the collaborative behaviour of the ant colony, whereas SPL is the shortest path length on SC (as a baseline for EPL)}
\label{table1}
\end{table}

\begin{table}
\centering
\begin{tabular}{| l | c | c | c | c |}
\hline
& \multicolumn{4}{c|}{\textbf{Arrival Measurements}} \tabularnewline
\hline
\textbf{Task } & \textbf{SC} & $\textbf{SC}_{\textbf{\scriptsize intra}}^{\textbf{\scriptsize rand}}$ & $\textbf{SC}_{\textbf{\scriptsize whole}}^{\textbf{\scriptsize rand}}$ & \textbf{SC}\\
&  \textbf{(AR)} & \textbf{(AR)} & \textbf{(AR)}  & \textbf{(MF)} \\
\hline
REST & 0.35 & 0.13 & -0.03 & -0.0043 \\
\hline
LANGUAGE & 0.42 & 0.17 & -0.03 & 0.0123 \\
\hline
EMOTION & 0.42 & 0.16 & -0.01 & 0.0109 \\
\hline
GAMBLING & 0.41 & 0.16 & 0.02 & 0.0054 \\
\hline
MOTOR & 0.48 & 0.20 & -0.02 & 0.0290 \\
\hline
RELATIONAL & 0.41 & 0.16 & 0.01 & -8 $\times$ $10^{-5}$ \\
\hline
SOCIAL & 0.38 & 0.15 & -0.01 & -0.0011\\
\hline
WM & 0.44 & 0.18 & -0.02 & 0.0078 \\
\hline
\end{tabular}
\vspace{6pt}
\caption{Pearson correlation coefficients between arrival measurements on SC and task-based FCs. Arrival measurements were obtained from SC and from two null models derived from SC, namely $SC_{intra}^{rand}$, and $SC_{whole}^{rand}$. AR is the arrival rate calculated from the collaborative behaviour of the ant colony, whereas MF is the maximum feasible flow between node pairs (baseline for AR).}
\label{table2}
\end{table}

\begin{table}
\centering
\begin{tabular}{| l | c | c | c | c |}
\hline
& \multicolumn{4}{c|}{\textbf{Combined Predictors}} \tabularnewline
\hline
\textbf{Task } & \textbf{SC} & $\textbf{SC}_{\textbf{\scriptsize intra}}^{\textbf{\scriptsize rand}}$ & $\textbf{SC}_{\textbf{\scriptsize whole}}^{\textbf{\scriptsize rand}}$ & \textbf{SC}\\
&  \textbf{(EPL,AR)} & \textbf{(EPL,AR)} & \textbf{(EPL,AR)}  & \textbf{(SPL,MF)}  \\
\hline
REST &0.14 & 0.02 & 0.0001 & 0.03  \\
\hline
LANGUAGE & 0.22 & 0.04 & 0.0001 & 0.10  \\
\hline
EMOTION & 0.21 & 0.04 &  0.0004 & 0.11 \\
\hline
GAMBLING & 0.20 & 0.03 & 0.0004 & 0.11 \\
\hline
MOTOR & 0.28 & 0.05 & 0.0010 & 0.14  \\
\hline
RELATIONAL & 0.18 & 0.03 & 0.0003 & 0.11  \\
\hline
SOCIAL & 0.17 & 0.03 & 0.0005 & 0.08 \\
\hline
WM & 0.23 & 0.04 & 0.0007 & 0.13 \\
\hline
\end{tabular}
\vspace{6pt}
\caption{Multi-linear models using path-length and arrival measurements as predictors of FC for different tasks. Values indicate explained variance. SC and two subsequent null models, $SC_{intra}^{rand}$, and $SC_{whole}^{rand}$ are evaluated. SPL and MF are evaluated on SC as baseline models for EPL and AR respectively.}
\label{table3}
\end{table}

\textit{Tables \ref{table1}} shows the Pearson correlation coefficients between the path length measurements and task-based FCs. The path length measurements were calculated on SC and two null models, $SC_{intra}^{rand}$ (intra-hemispheric randomization) and $SC_{whole}^{rand}$ (whole-brain randomization, see \textit{Section 2.5} for details). The table also reports the associations between FCs and shortest path length (SPL) as a baseline for EPL. Note that, as expected, EPL calculated on SC is negatively correlated with all task and resting-state FCs. This indicates that the longer the EPL for a node pair, the less functionally coupled it is. Also, for all tasks and resting-state FCs, the EPL computed on SC outperforms EPL on the two null models as well as SPL on SC.\\

Analogously, \textit{Tables \ref{table2}} shows the Pearson correlation coefficients between the arrival measurements and task-based FCs. The arrival measurements were calculated on SC and two null models, $SC_{intra}^{rand}$ (intra-hemispheric randomization) and $SC_{whole}^{rand}$ (whole-brain randomization, see \textit{Section 2.5} for details). The table also details the associations between FCs and max flow (MF) \cite{harris1955fundamentals} as a baseline for AR. Note that, as expected, AR calculated on SC is positively correlated with all task FCs. This indicates that the higher the AR for a node pair, the higher the functional coupling between them. Also, for all tasks and resting-state FCs, the AR computed on SC outperforms AR on the two null models as well as MF on SC.\\

In order to evaluate the joint predictive capacity of EPL and AR, we conducted multilinear regression analysis for every $\alpha$-$\beta$ configuration. \textit{Figure \ref{fig_regstats}.A} and \textit{B} show the explained variance ($R^2$) for all configurations for the resting-state and motor-task FCs respectively. \textit{Figure \ref{fig_regstats}.C} and \textit{D} are the scatter plots between the predicted and observed FCs for resting-state and motor-task respectively for the optimal configurations. \textit{Table \ref{table3}} summarizes the $R^2$ values for all FCs when the underlying topology is SC, $SC_{intra}^{rand}$, and $SC_{whole}^{rand}$. The predictive capacity of the EPL and AR when using the SC null models is negligible. Also observe that EPL and AR calculated on SC outperform the baseline predictors, SPL and MF.

\begin{figure}
	\includegraphics[width=\textwidth]{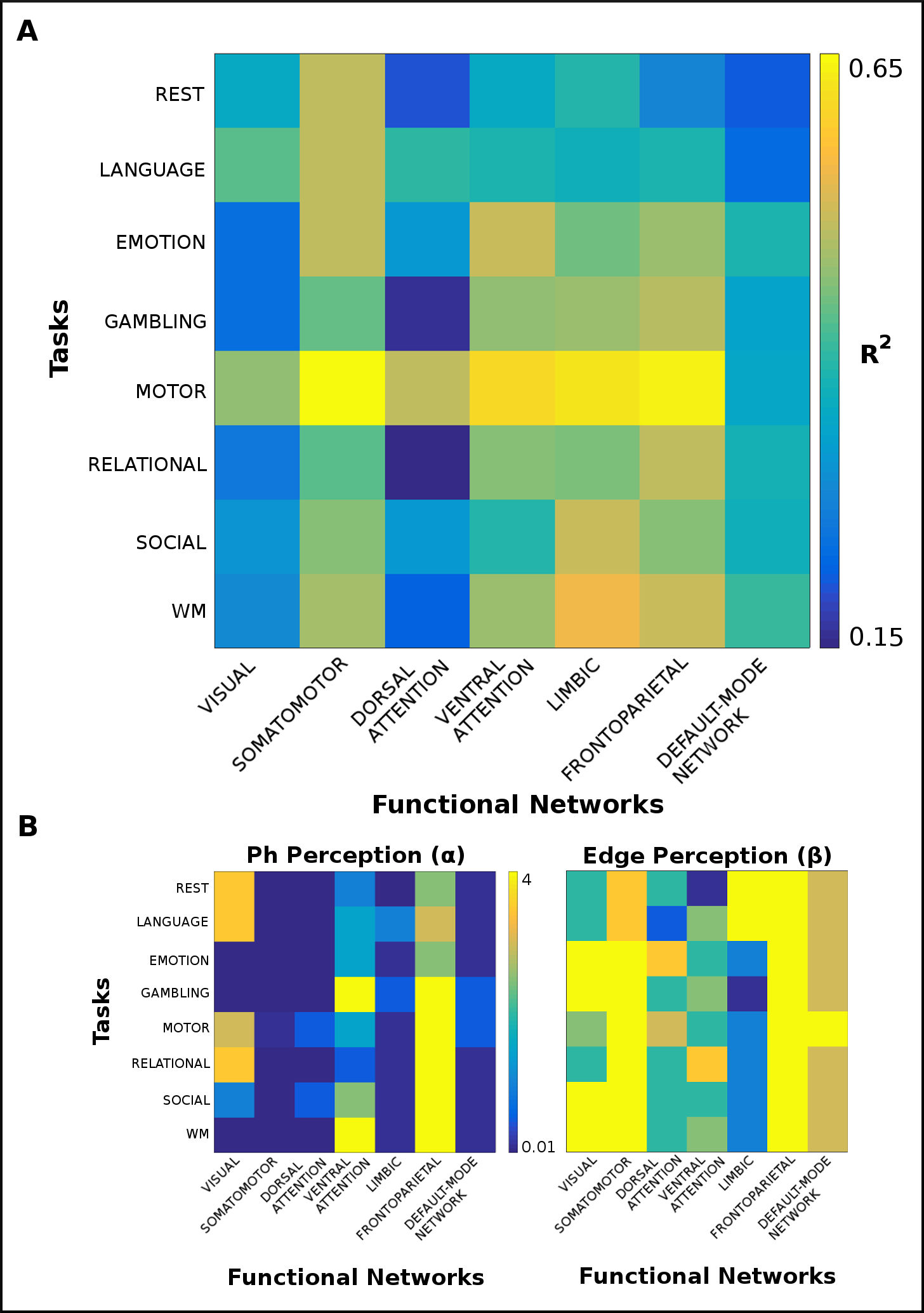}
    \caption{\textit{A.} Optimal configurations of ant-colony parameters, \textit{ph} and \textit{edge perception}. Each entry denotes the optimal configuration (highest $R^2$) for a task and a functional network (FN). Note that for each FN, only the node pairs involving brain regions of that FN are considered.}
    \label{fig_task_FN}
\end{figure}

\subsection{Associations within Functional Networks}

Lastly, we further tested the joint predictive capabilities of EPL and AR within the 7 brain functional networks (FN) as defined by Yeo et al. \cite{yeo2011organization}. This was achieved by carrying out multilinear regression analysis across all $\alpha$-$\beta$ configurations for node pairs within the 7 FNs (i.e., visual, somatomotor, dorsal attention, ventral attention, limbic, frontoparietal, and default-mode network). For each FN-FC combination, the highest explained variance ($R^2$) and its corresponding $\alpha$-$\beta$ configuration was saved. \textit{Figure \ref{fig_task_FN}.A} shows maximum $R^2$ values reached for each FC-FN combination, while \textit{Figure \ref{fig_task_FN}.B} shows the corresponding $\alpha$ and $\beta$ values.\\

Observe that Motor-task has the highest $R^2$ within frontoparietal FN, followed closely by somatomotor FN. Even though these $R^2$ values are very close, the big difference between them is the $\alpha$-$\beta$ configuration for which they are observed. Note that both $\alpha$ and $\beta$ are above 1 for frontoparietal region, while for somatomotor the configuration is $\alpha$ below 1 and $\beta$ above 1. Referring to \textit{Table \ref{table_alpha_beta}}, we can see that a highly collaborative regime and communication through the most prominent edges is essential for motor task within frontoparietal FN. On the other hand, within somatomotor FN, the ant colony regime is not collaborative but the communication takes place through the prominent edges in the network. This suggests that, for a single task, the communication within brain regions imitates different regimes of the ant colony algorithm within different FNs. Similar phenomena can be observed for other FC-FN combinations as well.

\section{Discussion}

There have been several studies in the recent past focused on a better understanding of the communication mechanisms of the human brain \cite{avena2018communication}. The present paper delves into this topic by proposing a framework inspired from the collaborative foraging behaviour of a colony of ants in order to simulate communication as a spreading phenomenon on the top of an underlaying complex network. This framework allowed characterization of source-target communications, not as a single estimate through a single (shortest) path but through path ensembles whose identification is sensitive to the pheromone-based activity of the ants. From \textit{Tables \ref{table1}, \ref{table2}}, and \textit{\ref{table3}} we can observe that the associations between the different FCs and the two measures that we have defined, EPL and AR, are significant when the algorithm is run on SC as compared to the two different null models and the baseline models in terms of SPL and MF.\\

Additionally, it was found that these associations are higher within certain functional networks for specific FCs (see \textit{Section 3.3} for details). This result in particular might be an indication that when a person is performing a task, a specific set of brain regions is more active than the other parts of the brain and that this activity is simulated well by the ant colony algorithm. Further, the communication regimes used by the ants in terms of $\alpha$ and $\beta$ are different for different FC-FN combinations. This suggests that the communication dynamic within different FNs varies when a person is performing different tasks. \\

An added value of the method proposed in this paper is that it allows for simple parametrization of any system between two layers,  its structural and functional sides, using just two parameters, namely \textit{pheromone perception} and \textit{edge perception}. Another point to note is that this framework obtains, in a data-driven fashion based on simulations, the path ensembles representing the most important communication pathways between each source and each target, as opposed to fixing the number of paths \cite{avena2017path} as a constant value for all path ensembles or fixing the number of steps \cite{fallani2011multiple}. In consequence, the presented framework allows for communication between different sources and targets to have different number of paths involved depending on the topology and the dynamics of the ant-colony as defined by \textit{pheromone perception} and \textit{edge perception}. The impact of these two exponents on the behavior of the ant colony and on the communication regime is summarized in \textit{Table \ref{table_alpha_beta}}. The presented framework allows for very different communication regimes occurring under the same topology, from independent random walkers that perceive the network as binary to highly collaborative walkers biased towards using either the \textit{main roads} or the \textit{side roads}.\\

The ant colony-inspired algorithm presented here as a dynamical model on top of a network topology is a framework that may be used as a testbed for evaluating different communication scenarios. Previously, Mi\u{s}i\'{c} et al. \cite{mivsic2015cooperative} have used a cascade spreading model to study the network properties of the human brain that facilitates spreading of signal through any possible path. In order to do so, they activated two or more \textit{seed nodes} in the SC and studied how the perturbation spreads through the network in a collaborative and also in a competitive manner. The primary difference between our approach to modeling communication in the brain vs. Mi\u{s}i\'{c} et al.'s approach is the use of pheromones as a means of indirect communication between the ants. Another difference in the two approaches is that while Mi\u{s}i\'{c} et al. do not allow a node that has already received a signal once to be perturbed by the same or another signal again, the ant colony simulations allow for multiple uses of the same paths. Indeed, this is how the ants strengthen certain paths more than the others in their search for the target node.\\

One of the biggest limitations of this study is the computational power and time required to run the ant colony simulations on the brain network. This limitation has restricted the simulations in terms of number of ants, values of pheromone amount explored, and the brain parcellation used. It has also prevented the use of individual connectomes, hence preventing us from assessing inter-subject differences that should be explored in future work.\\

Future work shall be focused on exploring the inter-subject differences in connectomes through the algorithm proposed in this paper. As this paper used data from healthy subjects, another avenue is to study the behaviour of the ant colony on the connectomes of patients of neurodegenerative disorders, such as Alzheimer's or multiple sclerosis. As this framework is source-target oriented, it could also be linked with experiments where the concept of a source is very well-defined, such as Transcranial Magnetic Stimulation (TMS) \cite{pascual2011characterizing}\cite{amico2017tracking}. This framework can also be used in order to assess other systems, such as social or road networks.\\

The framework presented here combines a complex network topology tested by an ant-colony algorithm that, by means of two perception exponents, namely \textit{ph} and \textit{edge perception}, allows to simulate different communication regimes and to capture the most important path ensembles involved on the communication of each pair of source and target nodes. This framework has shown evidence of being able to establish associations between SC and FC when subjects are in different cognitive states as they are performing different tasks. This methodology allows for compression of the communicability happening to a reduced two-parameter space. We have presented important foundations on how these parameters mimic different communication regimes that might better explain different functional states.

\section*{Acknowledgments}

Data were provided [in part] by the Human Connectome Project, WU-Minn Consortium (Principal Investigators: David Van Essen and Kamil Ugurbil; 1U54MH091657) funded by the 16 NIH Institutes and Centers that support the NIH Blueprint for Neuroscience Research; and by the McDonnell Center for Systems Neuroscience at Washington University. This work was partially supported by  IHR01EB022574,by NIH R01MH108467, and by the Indiana Clinical and Translational Sciences Institute (UL1TR001108) from the NIH, National Center for Advancing Translational Sciences, Clinical and Translational Sciences Award. We would like to thank Dr. Olaf Sporns for insightful discussions.

\nolinenumbers

\bibliography{bibliography2}

\bibliographystyle{abbrv}

\end{document}